\input{aipcheck}

\documentclass[
    ,final            
  ]
  {aipproc}

\layoutstyle{6x9}

\def\lan{\langle}
\def\ran{\rangle}

\def\beq{\begin{equation}}
\def\ee{\end{equation}}
\def\eeq{\end{equation}}

\def\bfig{\begin{figure}}
\def\efig{\end{figure}}
\def\bea{\begin{eqnarray}}
\def\beann{\begin{eqnarray*}}
\def\eea{\end{eqnarray}}
\def\eeann{\end{eqnarray*}}

\def\raw{\rightarrow}

\usepackage{graphicx}

\usepackage{epsfig}

\begin{document}

\title{Glueball decay in the Fock-Tani formalism}

\classification{12.39.Mk; 12.39.Pn }
\keywords      {Glueball decay; Fock-Tani Formalism}

\author{Mario L. L. da Silva}{
  address={Instituto de F\'{\i}sica, Universidade Federal do Rio Grande 
do Sul, Caixa Postal 15051, CEP 91501-970, Porto Alegre, RS, Brazil.}
}

\author{Daniel T. da Silva}{
  address={Instituto de F\'{\i}sica, Universidade Federal do Rio Grande 
do Sul, Caixa Postal 15051, CEP 91501-970, Porto Alegre, RS, Brazil.}
}

\author{Dimiter Hadjimichef}{
  address={Departamento de F\'{\i}sica, Instituto de F\'{\i}sica 
e Matem\'atica, Universidade Federal de Pelotas, Campus Universit\'ario, 
CEP 96010-900, Pelotas, RS, Brazil.}
  ,altaddress={Instituto de F\'{\i}sica, Universidade Federal do Rio Grande 
do Sul, Caixa Postal 15051, CEP 91501-970, Porto Alegre, RS, Brazil.} 
}

\author{Cesar A. Z. Vasconcellos}{
  address={Instituto de F\'{\i}sica, Universidade Federal do Rio Grande 
do Sul, Caixa Postal 15051, CEP 91501-970, Porto Alegre, RS, Brazil.}
}

\begin{abstract}
We investigate the two-meson decay modes for $f_0(1500)$. 
In this calculation we consider this resonance
as a glueball.    The Fock-Tani formalism is introduced to calculate the
decay width.

\end{abstract}

\maketitle


\section{Introduction}
The gluon self-coupling in QCD opens the possibility of   existing  bound states of pure 
gauge fields known as glueballs. Even though theoretically acceptable, the question still 
remains unanswered: do bound states of gluons actually exist?  
Glueballs are predicted by many models and by lattice
calculations. 
In experiments glueballs are supposed to be produced in gluon-rich environments. 
The most important reactions to study gluonic degrees of freedom are radiative $J/\psi$ decays, 
central productions processes and antiproton-proton annihilation.

Numerous technical difficulties have so far been pre\-sent
in our understanding of their properties in experiments, largely because glueball states
can mix strongly with near\-by $q\bar{q}$ resonances \cite{amsler1},\cite{amsler2}. 
However recent experimental and 
lattice studies of $0^{++}$, $2^{++}$ and $0^{-+}$ glueballs seem to be convergent.

In the present we shall apply the Fock-Tani formalism \cite{annals} to glueball decay. In particular the 
resonance $f_0(1500)$ shall be considered. On 
theoretical grounds, a simple potential model with massive constituent gluons, 
namely the model of Cornwall and Soni \cite{cs1},\cite{cs2} has received attention
recently \cite{cs3},\cite{cs4} for spectroscopic calculations. The results obtained are 
consistent with lattice and experiment.

\section{The Fock-Tani formalism}
Now let us to apply the Fock-Tani formalism in the microscopic 
Hamiltonian to obtain an effective Hamiltonian. In the Fock-Tani
formalism we can write the glueball and the meson creation 
operators in the following form
\begin{eqnarray}
  G_{\alpha}^{\dagger}=\frac{1}{\sqrt{2}}\Phi_{\alpha}^{\mu \nu}
  a_{\mu}^{\dagger}a_{\nu}^{\dagger}.\,\,\,\,;\,\,\,\,
  M_{\alpha}^{\dagger}=\Psi_{\alpha}^{\mu \nu}
  q_{\mu}^{\dagger}\bar{q}_{\nu}^{\dagger}.
  \label{G}
\end{eqnarray}  
The gluon creation $a^{\dag}_{\nu}$ and annihilation $a_{\mu}$ operators  
obey the following commutation relations $ [a_{\mu},a_{\nu}]=0$ 
and $[a_{\mu},a_{\nu}^{\dagger}]=\delta_{\mu\nu}$.
While the quark creation $q^{\dag}_{\nu}$, annihilation $q_{\mu}$, the
antiquark creation $\bar{q}^{\dag}_{\nu}$ and annihilation $\bar{q}_{\mu}$
operators obey the following anticommutation relations 
$ \{q_{\mu},q_{\nu}\}=\{\bar{q}_{\mu},\bar{q}_{\nu}\}=\{q_{\mu},\bar{q}_{\nu}\} 
  =\{q_{\mu},\bar{q}_{\nu}^\dag\} = 0 $ and $ \{q_{\mu},q_{\nu}^{\dagger}\}
=\{\bar{q}_{\mu},\bar{q}_{\nu}^{\dagger}\}=\delta_{\mu\nu}$.
In (\ref{G}) $\,\Phi_{\alpha}^{\mu\nu}$ and $\,\Psi_{\alpha}^{\mu\nu}$ are 
the bound-state wave-functions for two-gluons and two-quarks respectively.
The composite glueball and meson operators satisfy non-canonical commutation 
relations
\begin{eqnarray}
  & &[G_{\alpha},G_{\beta}]=0\,\,\,\,\,;\,\,\,\, 
  [G_{\alpha},G_{\beta}^{\dagger}]
  =\delta_{\alpha\beta}+\Delta_{\alpha\beta}\nonumber\\
  & &[M_{\alpha},M_{\beta}]=0\,\,\,\,\,;\,\,\,\, 
  [M_{\alpha},M_{\beta}^{\dagger}]
  =\delta_{\alpha\beta}+\Delta_{\alpha\beta}
\end{eqnarray}
The  ``ideal particles'' which obey canonical relations
\begin{eqnarray}
  & &[g_{\alpha},g_{\beta}]=0\,\,\,\,\,;\,\,\,\, [g_{\alpha},g_{\beta}^{\dagger}]
  =\delta_{\alpha\beta}\nonumber\\
  & &[m_{\alpha},m_{\beta}]=0\,\,\,\,\,;\,\,\,\, [m_{\alpha},m_{\beta}^{\dagger}]
  =\delta_{\alpha\beta}.
\end{eqnarray}
This way one can transform the composite state $|\alpha\rangle$ into an ideal state
$|\alpha\,)$, in the glueball case for example we have
\begin{eqnarray*}
  |\alpha\,)=   U^{-1}(-\frac{\pi}{2})\,G_{\alpha}^{\dagger}
  \,|0\rangle=g_{\alpha}^{\dagger}\,| 0\rangle
\end{eqnarray*}
where $ U=\exp({tF})$ and $F$ is the generator of the glueball transformation given by
\begin{eqnarray}
  F=\sum_\alpha g_{\alpha}^{\dagger}\tilde{G}_{\alpha}
  - \tilde{G}_{\alpha}^{\dagger}g_{\alpha}
  \label{F}
\end{eqnarray}
with
\begin{eqnarray*}
  \tilde{G}_{\alpha}=G_{\alpha}-\frac{1}{2}\Delta_{\alpha\beta}G_{\beta}
  -\frac{1}{2}G_{\beta}^{\dagger}[\Delta_{\beta\gamma},G_{\alpha}]G_{\gamma}.
\end{eqnarray*}
In order to obtain the effective potential one has to use (\ref{F}) in a 
set of Heisenberg-like equations for the basic operators $g,\tilde{G},a$
\begin{eqnarray*}
  \frac{dg_{\alpha}(t)}{dt}=[g_{\alpha},F]=\tilde{G}_{\alpha}\,\,\,\,\,;\,\,\,\,\,
  \frac{d\tilde{G}_{\alpha}(t)}{dt}=[\tilde{G}_{\alpha}(t),F]=-g_{\alpha}\,.
\end{eqnarray*}
The simplicity of these equations are not present in the equations for $a$
\begin{eqnarray*}
  \frac{da_{\mu}(t)}{dt}=[a_{\mu},F]=&-&\sqrt{2}\Phi_{\beta}^{\mu\nu}a_{\nu}^{\dagger}g_{\beta}
  +\frac{\sqrt{2}}{2}\Phi_{\beta}^{\mu\nu}a_{\nu}^{\dagger}\Delta_{\beta\alpha}g_{\beta}\\
  &+&\Phi_{\alpha}^{\star\mu\gamma}\Phi_{\beta}^{\gamma\mu^{'}}
  (G_{\beta}^{\dagger}a_{\mu^{'}}g_{\beta}-g_{\beta}^{\dagger}a_{\mu^{'}}G_{\beta})\\
  &-&\sqrt{2}(\Phi_{\alpha}^{\mu\rho^{'}}\Phi_{\rho}^{\mu^{'}\gamma^{'}}
  \Phi_{\gamma}^{\star\gamma^{'}\rho^{'}}
  +\Phi_{\alpha}^{\mu^{'}\rho^{'}}\Phi_{\rho}^{\mu\gamma^{'}}
  \Phi_{\gamma}^{\star\gamma^{'}\rho^{'}})G_{\gamma}^{\dagger}a_{\mu^{'}}^{\dagger}
  G_{\beta}g_{\beta}.
\end{eqnarray*}
The solution for these equation can be found order by order in the wave 
functions. For zero order one has $a_{\mu}^{(0)}=a_{\mu}$,
$ g_{\alpha}^{(0)}(t)=G_{\alpha}\sin{t}+g_{\alpha}\cos{t}$ and
$  G_{\beta}^{(0)}(t) =G_{\beta}\cos{t}-g_{\beta}\sin{t}$.
In the first order $g_{\alpha}^{(1)}=0,\,\,\,G_{\beta}^{(1)}=0$ and
$a_{\mu}^{(1)}(t)=\sqrt{2}\Phi_{\beta}^{\mu\nu}a_{\nu}^{\dagger}g_{\beta}$.
If we repeat a similar calculation for mesons let us to obtain the
following equations solution: $q_{\mu}^{(0)}=q_{\mu}$, 
$\bar{q}_{\mu}^{(0)}=\bar{q}_{\mu}$,
$  q_{\mu}^{(1)}(t)=\Psi_{\beta}^{\mu\nu}\bar{q}_{\nu}^{\dagger}m_{\beta}$ and
$  \bar{q}_{\mu}^{(1)}(t)= -\Psi_{\beta}^{\mu\nu}q_{\nu}^{\dagger}m_{\beta}$.

\section{The Microscopic Model}

The microscopic model adopted here must contain explicit quark and
gluon degrees of freedom, so we obtain a microscopic 
Hamiltonian of the following form 
\begin{eqnarray}
  H = g^2 \int d^3x d^3y \Psi^{\dag}(\vec{x}) \gamma^0 \gamma^i A_i^a
  (\vec{x}) \frac{\lambda_{\mu\nu}^a}{2} \Psi(\vec{x}) \Psi^{\dag}(\vec{y})
  \gamma^0 \gamma^j A_j^b (\vec{y}) \frac{\lambda_{\sigma\rho}^b}{2}
  \Psi(\vec{y})
\end{eqnarray}
Where the quark and the gluon fields are respectively \cite{swanson}
\begin{eqnarray}
  \Psi(\vec{x}) = \sum_s \int \frac{d^3k}{(2\pi)^3} [u(\vec{k},s)
  q(\vec{k},s) + v(-\vec{k},s) \bar{q}^\dag(-\vec{k},s)]
  e^{i\vec{k}\cdot\vec{x}}
\end{eqnarray}
and
\begin{eqnarray}
  A_i^a (\vec{x}) = \int \frac{d^3k}{(2\pi)^3}
  \frac{1}{\sqrt{2\omega_{\vec{k}}}} [a_i^a(\vec{k})
    + {a_i^a}^\dag (-\vec{k})] e^{i\vec{k}\cdot\vec{x}}
\end{eqnarray}
We choose this Hamiltonian due to its form that allow to obtain
a operators structure of this type
$ q^\dag \bar{q}^\dag q^\dag \bar{q}^\dag a a$

\section{The Fock-Tani formalism application}

\begin{figure}[ht]
\scalebox{.8}{\includegraphics*[80pt,550pt][270pt,670pt]{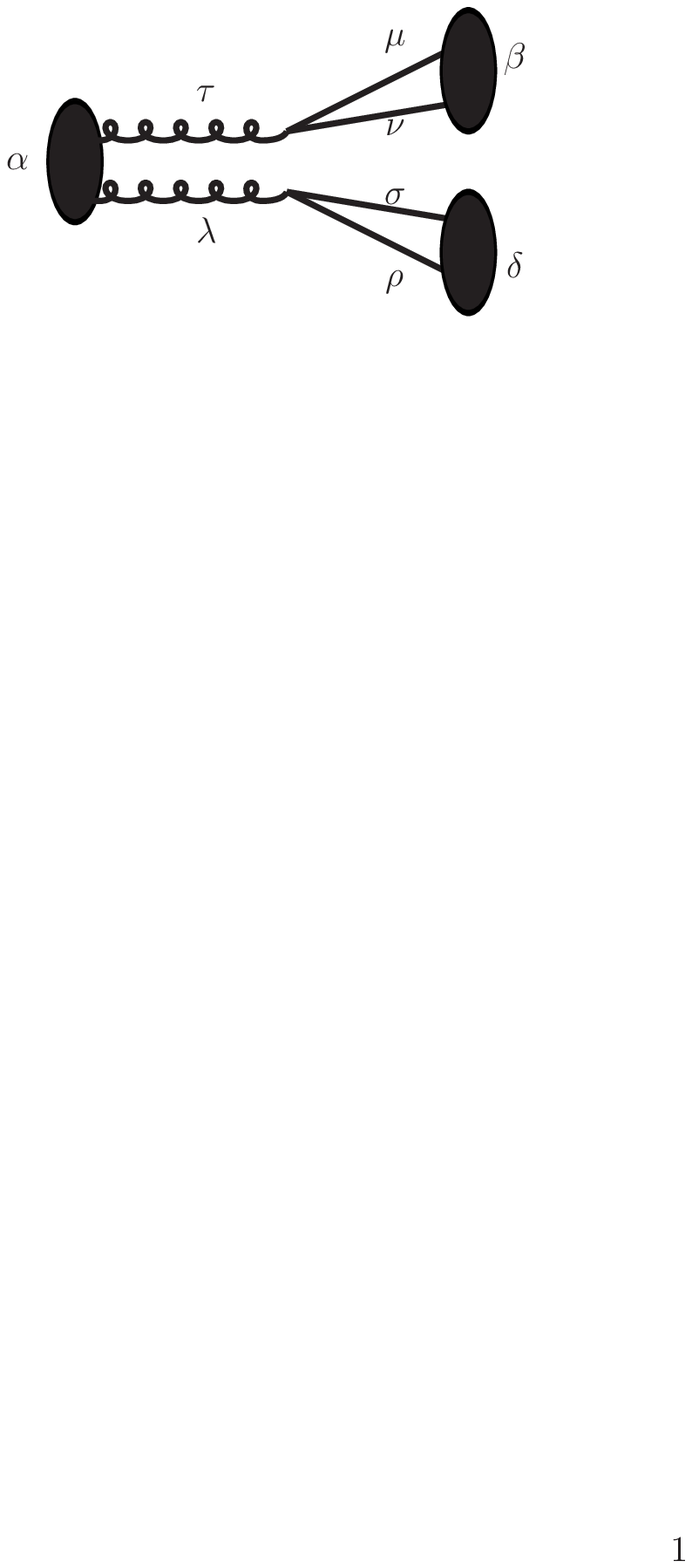}}
    \scalebox{.8}{\includegraphics*[80pt,550pt][270pt,670pt]{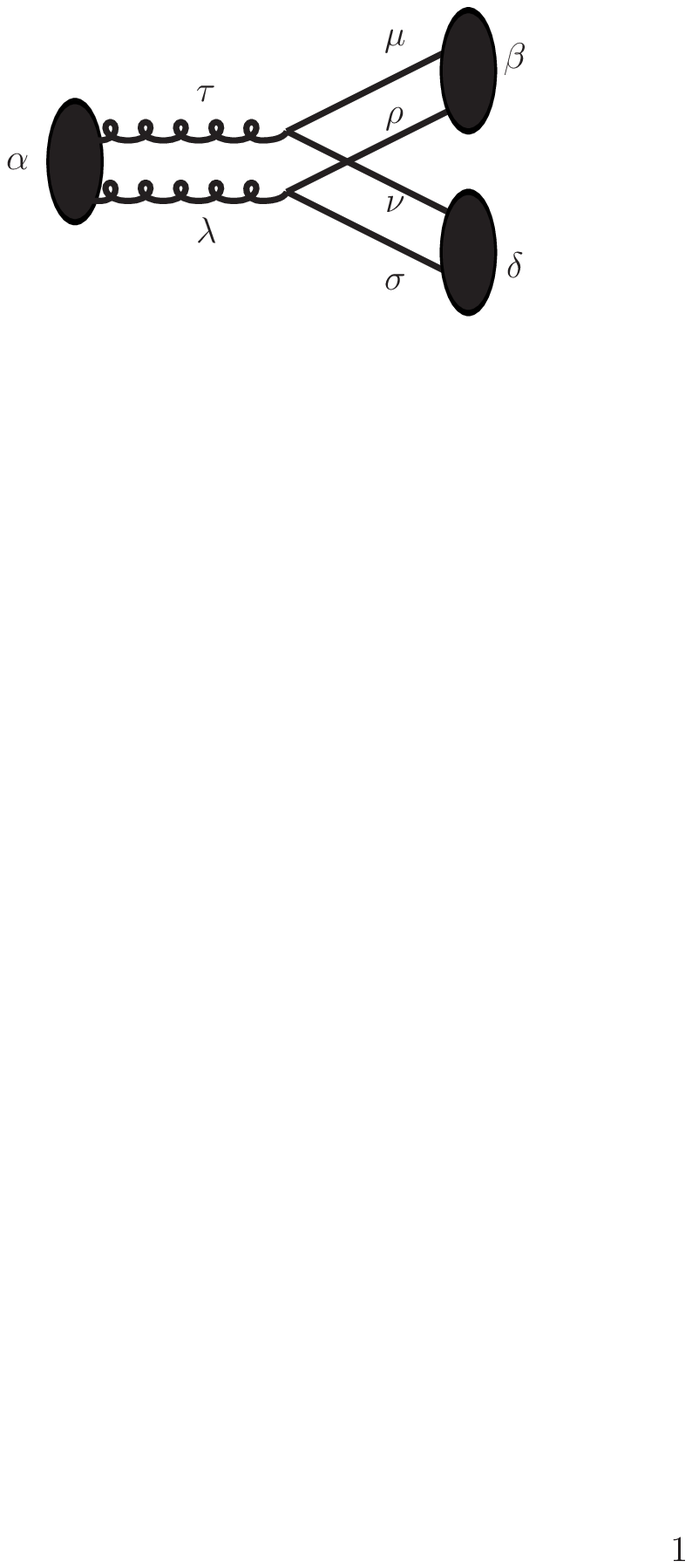}}
      \caption{Diagrams for  glueball decay  } 
      \label{diagrama}
\end{figure}

Now we are going to apply the Fock-Tani formalism to the microscopic
Hamiltonian
\begin{eqnarray}
  H_{FT} = U^{-1} H U
\end{eqnarray}
which gives rise to an effective interaction $H_{FT}$. 
To find this  Hamiltonian we have to calculate the transformed operators for
quarks and gluons by a technique known as 
{\it the equation of motion technique}. The resulting $H_{FT}$ for the glueball
decay $G\raw mm$ is represented by two diagrams which appear in Fig.~(\ref{diagrama}).

Analyzing these diagrams, of  Fig.~(\ref{diagrama}), it is clear
that in the first one there is no color conservation. 
The glueball's wave-function $\Phi$ is written as a product
\bea
\Phi_{\alpha}^{\mu\nu}=
\chi_{A_{\alpha}}^{s_{\mu}s_{\nu}}\,
{\cal C}^{c_{\mu}c_{\nu}}\,
\Phi_{\vec{P}_{\alpha} }^{\,\vec{p}_{\mu}\vec{p}_{\nu}},
\label{wf}
\eea
$\chi_{A_{\alpha}}^{s_{\mu}s_{\nu}}$ 
is the spin contribution, with $A_{\alpha}\equiv \{S_{\alpha},S^{3}_{\alpha}\} $,
where $S_{\alpha}$ is the glueball's  total spin index  and $S^{3}_{\alpha} $ 
the index of the spin's third component; ${\cal C}^{c_{\mu}c_{\nu}}$ is the color 
component  given by $\frac{1}{\sqrt{8}} \,\delta^{c_{\mu}c_{\nu}}$
and the spatial wave-function is
\bea
\Phi_{\vec{P}_{\alpha}}^{\vec{p}_{\mu}\vec{p}_{\nu}}=
\delta^{(3)}(\vec{P}_{\alpha}-\vec{p}_{\mu}-\vec{p}_{\nu})\,
{\left(\frac{1}{\pi
b^2}\right)}^{\frac{3}{4}}e^{-\frac{1}{8\beta^{2}}{
\left(\vec{p}_{\mu}-\vec{p}_{\nu}\right)}^{2}}.
\label{phi-7}
\eea
The expectation value of $r^2$ gives a relation between 
the $rms$ radius $r_0$ and $\beta$ of the form  $\beta=\sqrt{1.5}/r_0$.
The meson wave function $\Psi$ is similar with  parameter $b$ replacing $\beta$.
To determine the decay rate, we evaluate the matrix element between the states 
$|i\ran=g^{\dag}_{\alpha}|0\ran$ and $|f\ran=m^{\dag}_{\beta} m^{\dag}_{\gamma}   |0\ran$
which is of the form
$\lan f \mid H_{FT} \mid  i \ran= \delta (\vec{p}_{\alpha}-\vec{p}_{\beta}-\vec{p}_{\gamma})
h_{hi}$.
The $h_{fi}$ decay amplitude can be combined with a relativistic phase space to give
the differential decay rate \cite{barnes}
\bea
\frac{d\Gamma_{\alpha\raw \beta\gamma}}{d\Omega}=2\pi\, \frac{PE_{\beta}E_{\gamma} }{ M_{\alpha}}
\,\,|h_{fi}|^2
\eea
After several manipulations we obtain the following result
\begin{eqnarray}
  \lan f \mid H_{FT} \mid  i \ran = 
  \frac{8\alpha_s}{3\pi} \left(\frac{1}{\pi b^2}
  \right)^{3/4} \int dq \, \frac{q^2}{\sqrt{q^2 + m_g^2}} 
  \left(1  - \frac{q^2}{4m_q^2} - \frac{q^2}{4m_{s}^2}      \right)
  e^{-(\frac{1}{2b^2} + \frac{1}{4\beta^2}){q}^2}
\end{eqnarray}
Finally one can write the decay amplitude for the $f_0$ into two
mesons
\bea
  \Gamma_{f_0\rightarrow M_1 M_2} =  \frac{512\alpha_s^2}{9} \,\,
  \frac{P\,E_{M_1}E_{M_2}}{M_{f_0}}
\left(\frac{1}{\pi b^2}
  \right)^{3/2}\, {\cal I}^{\,2}
\eea
where 
\bea
{\cal I}=\int dq \, \frac{q^2}{\sqrt{q^2 + m_g^2}} 
  \left(1  - \frac{q^2}{4m_{q}^2} - \frac{q^2}{4m_{s}^2}      \right)
  e^{-(\frac{1}{2b^2} + \frac{1}{4\beta^2}){q}^2}
\eea
with $m_q$ the $u$ and $d$ quark mass and $m_{s}$ the mass of the $s$ quark.
The decays that are studied are for the following processes $f\raw \pi\pi$,
$f\raw K\bar{K}$ and $f\raw \eta\eta$. The parameters used are $b=0.34$ GeV,
$m_q=0.33$, $m_q/m_s=0.6$, $\alpha_s=0.6$. 

\begin{figure}[ht]
  \scalebox{.4}{\includegraphics*{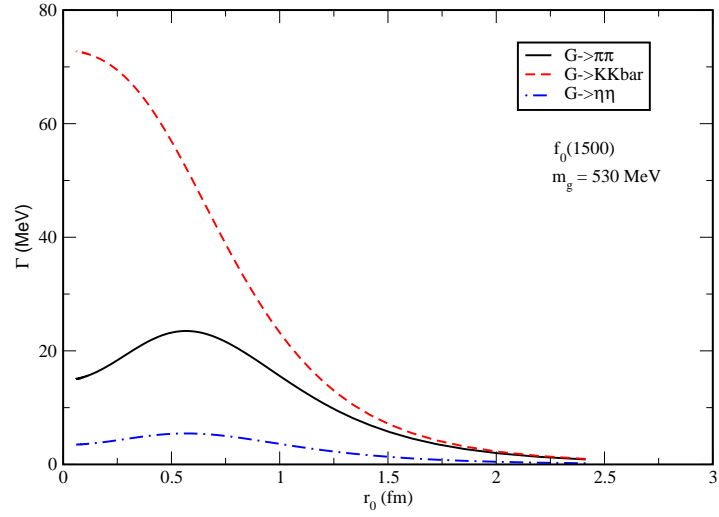}}
  \caption{Decay width for $f_0(1500)$} 
  \label{fig}
\end{figure}

\section{Conclusions}
The Fock-Tani formalism is proven appropriate not only for hadron scattering but for decay.
The example decay process $f_{0}(1500)\rightarrow \pi\pi$; $K\bar{K}$ and $\eta\eta$ in the 
Fock-Tani formalism is studied.
The same procedure can be used for other $f_{0}\left(M\right)$ and for heavier scalar mesons 
and compared with similar calculations which include mixtures.

\begin{theacknowledgments}
The authors acknowledges support from the Funda\c c\~ao de Amparo \`a Pesquisa 
do Estado do Rio Grande do Sul - FAPERGS. M.L.L.S. acknowledges support 
from the Conselho Nacional de Desenvolvimento Cient\'{\i}fico e Tecnol\'ogico
- CNPq. D.T.S. acknowledges support from the Coordena\c c\~ao de 
Aperfei\c coamento de Pessoal de N\'{\i}vel Superior - CAPES.
\end{theacknowledgments}



\bibliographystyle{aipproc}   

\bibliography{sample}

\IfFileExists{\jobname.bbl}{}
 {\typeout{}
  \typeout{******************************************}
  \typeout{** Please run "bibtex \jobname" to optain}
  \typeout{** the bibliography and then re-run LaTeX}
  \typeout{** twice to fix the references!}
  \typeout{******************************************}
  \typeout{}
 }

\end{document}